\documentclass[preprint,superscriptaddress]{revtex4-1}

\usepackage{graphicx}
\usepackage{epstopdf}
\usepackage[ansinew]{inputenc}
\usepackage{array}
\usepackage{color}
\usepackage{amsmath}
\usepackage{amsxtra}
\usepackage{amstext}
\usepackage{amssymb}
\usepackage{latexsym}
\usepackage{float}
\usepackage{bm}
\usepackage{dsfont}
\usepackage{textcomp}
\usepackage[T1]{fontenc}

\newcommand{\fref}[1]{Fig.~\ref{fig:#1}}

\begin{document}

\title{Ambipolar Landau levels and strong band-selective carrier interactions in monolayer WSe$_2$}
\author{Martin V. Gustafsson*}
\affiliation{Department of Chemistry, Columbia University, New York, NY, USA}
\affiliation{Department of Physics, Columbia University, New York, NY, USA}
\thanks{These authors contributed equally to this work.}
\author{Matthew Yankowitz*}
\affiliation{Department of Physics, Columbia University, New York, NY, USA}
\thanks{These authors contributed equally to this work.}
\author{Carlos Forsythe}
\affiliation{Department of Physics, Columbia University, New York, NY, USA}
\author{Daniel Rhodes}
\affiliation{Center for Integrated Science and Engineering, Columbia University, New York, NY, USA}
\author{Kenji Watanabe}
\affiliation{National Institute for Materials Science, 1-1 Namiki, Tsukuba 305-0044, Japan}
\author{Takashi Taniguchi}
\affiliation{National Institute for Materials Science, 1-1 Namiki, Tsukuba 305-0044, Japan}
\author{James Hone}
\affiliation{Department of Mechanical Engineering, Columbia University, New York, NY, USA}
\author{Xiaoyang Zhu}
\affiliation{Department of Chemistry, Columbia University, New York, NY, USA}
\author{Cory R. Dean}
\affiliation{Department of Physics, Columbia University, New York, NY, USA}
\date{\today}

\begin{abstract}
Monolayers (MLs) of transition metal dichalcogenides (TMDs) exhibit unusual electrical behavior under magnetic fields due to their intrinsic spin-orbit coupling and lack of inversion symmetry~\cite{Mak2010,Xiao2012,Li2013,Rose2013,Xu2014,Li2014,MacNeill2015,Srivastava2015,Aivazian2015,Stier2016,Wang2017,Wang2017b,Cui2015,Fallahazad2016,Movva2017}. While recent experiments have also identified the critical role of carrier interactions within these materials~\cite{Wang2017,Movva2017}, a complete mapping of the ambipolar Landau level (LL) sequence has remained elusive. Here, we use single-electron transistors~\cite{Wei1997,Martin2008} to perform LL spectroscopy in ML WSe$_2$, for the first time providing a comprehensive picture of the electronic structure of a ML TMD for both electrons and holes. We find that the LLs differ notably between the two bands, and follow a unique sequence in the valence band (VB) that is dominated by strong Zeeman effects. The Zeeman splitting in the VB is several times higher than the cyclotron energy, far exceeding the predictions of a single-particle model, and moreover tunes significantly with doping~\cite{Movva2017}. This implies exceptionally strong many-body interactions, and suggests that ML WSe$_2$ can serve as a host for new correlated-electron phenomena.
\end{abstract}

\maketitle

The semiconducting transition-metal dichalcogenides (TMDs) consist of stacked honeycomb lattices which, like graphene, can be exfoliated into monolayers (MLs)~\cite{Mak2010}. The ML bandgap is direct and degenerate at two unique points in the Brillouin zone, the K and K$^\prime$ valleys. ML TMDs lack inversion symmetry and exhibit strong spin-orbit coupling and, as a result, there is a sizable valley-dependent lifting of the spin degeneracy within each valley. This gives rise to a locking of the spin and valley degrees of freedom into one composite isospin (Fig. 1a)~\cite{Rose2013,Xu2014}. At high magnetic fields these unusual features become prominent, as the energy bands of the two-dimensional carriers break up into a series of degenerate Landau levels (LLs). The energy gaps between LLs are expected to be small in ML TMDs due to their large effective carrier masses~\cite{Mak2010,Fallahazad2016}. At the same time, the spin-orbit coupling and broken lattice symmetry give rise to Zeeman energy terms in addition to the magnetic coupling to the electron spin, causing the net Zeeman effect to be unusually strong. A single-particle model of the LLs in WSe$_2$ predicts a scenario where the Zeeman shift exceeds the LL separation in the valence band (VB), resulting in a large energy offset between LLs of opposite isospin (Fig. 1b). Additionally, the large effective mass suggests that many-body interactions can be strong even at high carrier densities, which can potentially alter the LL hierarchy beyond the already unusual predictions of the single-particle model~\cite{Wang2017,Movva2017}.

Some of the peculiar electronic phenomena in ML TMDs have been revealed with optical techniques~\cite{Li2014,MacNeill2015,Srivastava2015,Aivazian2015,Stier2016,Wang2017,Wang2017b} and by electrical transport~\cite{Cui2015,Fallahazad2016,Movva2017}. However, a direct mapping of the ambipolar LL structure is still lacking, primarily owing to poor material quality and difficulty in making transparent electrical contact for both carrier types~\cite{Allain2015}. We address both of these challenges by combining improved crystal growth with a local electrostatic probing technique that requires no steady-state current to flow through the TMD contacts. Our custom chalcogen-flux growth technique produces crystals with defect densities up to three orders of magnitude lower than commercial material (see Supplementary Information). Figs. 2a and b illustrate our measurement setup. We contact the exfoliated MLs with graphite sheets and encapsulate them between two flakes of hexagonal boron nitride resting on a graphite back gate~\cite{Wang2013}. On top of the stacked 2D materials, we deposit metallic single-electron transistors (SETs), which couple capacitively to the embedded ML. The SET is gated by variations in the local chemical potential $\mu$ of the TMD ML, providing a probe of the LL sequence and allowing us to determine the inter-LL energy gaps~\cite{Wei1997,Martin2008,Eisenstein1992}. The data presented in this paper were all acquired from the same sample, but we have also studied another sample from a different growth cycle which shows the same qualitative features (see Supplementary Fig. 2).

\begin{figure}
\centering
\includegraphics[width=8.8cm]{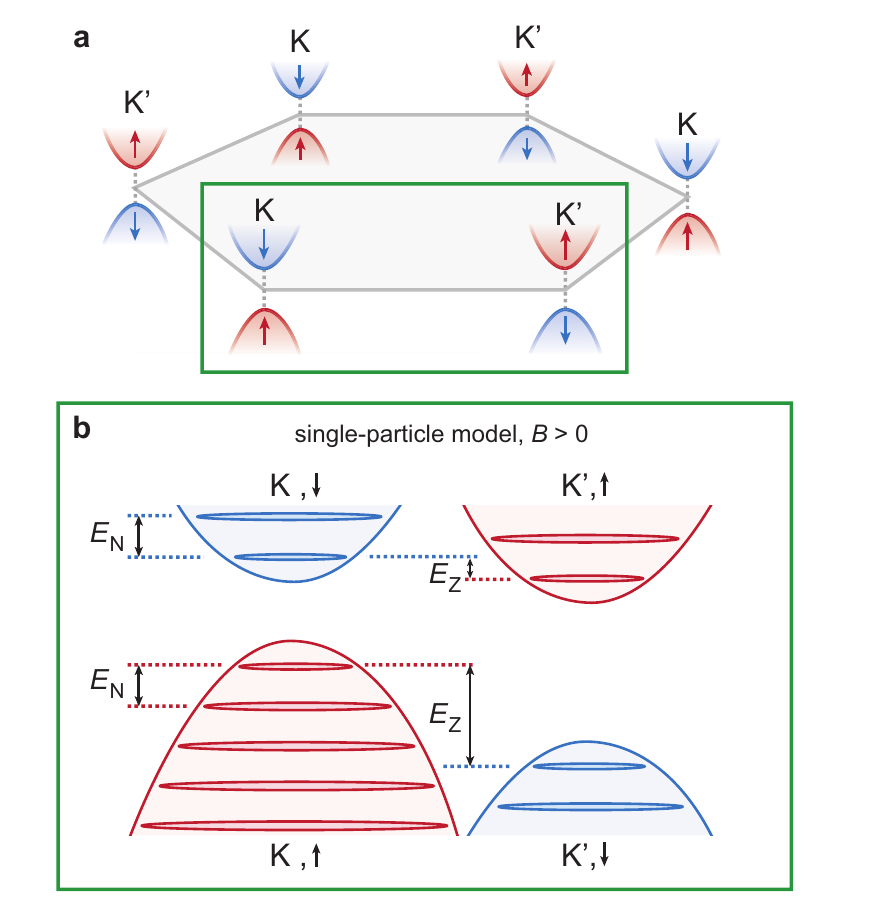} 
\caption{\textbf{| Monolayer WSe$_2$ in the quantum Hall regime.} \textbf{a}, Momentum-space illustration of the lowest energy bands in ML WSe$_2$. For each band, the valley index (K, K$^\prime$) is tied to a specific electron spin (up, red; down, blue). \textbf{b}, Anticipated LLs in WSe$_2$ in the absence of many-body interactions. In the VB, the Zeeman shift $E_Z$ exceeds the cyclotron energy $E_N$ by over a factor of 2, whereas it is much less pronounced in the CB. Only the lowest spin branch in each valley is shown.}
\end{figure}

We measure $\mu$ and the inverse compressibility $d \mu/d n$, where $n$ is the carrier density, as functions of back gate voltage $V_{BG}$ and magnetic field up to $B = 34.5$ T. Energy gaps between Landau levels are identifiable as maxima in $d\mu /dn$, which is inversely proportional to the denisty of states. Plotting $d\mu /dn$ versus $B$ and $n$ (Fig.~2c) reveals a characteristic ambipolar fan diagram. The positions of the gaps are well fit by the general relation $B=\frac{1}{\nu }\frac{n h}{e}$ in both the VB and CB, where $e$ is the electron charge, $h$ the Planck constant, and $\nu $ the filling factor ($\nu > 0$ in the CB  and $\nu < 0$ in the VB), allowing us to accurately identify the filling factor associated with each gap. By extrapolating the slopes of the LL gaps to $B = 0$, we find a separation in gate voltage of 2.7 V between the two bands, arising from the band gap of the material (see Supplementary Information). The higher spin-split bands are expected to be $\sim$ 500 meV and $\sim$ 30 meV removed from the lowest-energy VB and CB, respectively~\cite{Komider2013}, and as a result we only probe the lowest spin-split bands within our experimentally accessible density range (Fig.~1a).

The observation of quantum Hall features at all integer values confirms that, at least in the high field limit, all degeneracies have been lifted.  The ground state order associated with each state, however, depends sensitively on the relative size of the Zeeman splitting. We can write the Zeeman energy in terms of an effective Land\'e g-factor $g^*$ as $E_Z = 2\,|E_{Z,S} + E_{Z,B} + E_{Z,O}| = 2\,g^*\mu_B B$. Here $\mu_B$ is the Bohr magneton, and the three contributing Zeeman terms stem from magnetic coupling to the electron spin ($E_{Z,S}$), the valley-dependent Berry curvature ($E_{Z,B}$), and the orbital angular momenta from the tungsten nuclei ($E_{Z,O}$)~\cite{Aivazian2015}, respectively. The three contributions add with the same sign in the VB, resulting in $g^* \approx$ 5.5 in the absence of interactions. To determine the electronic properties of the material and test the single-particle predictions, it is thus essential to determine the magnitude of the Zeeman splitting. While this is conventionally done in measurements of electronic transport in a tilted magnetic field, this technique is not effective for ML TMDs due to a locking of the spins perpendicular to the 2D plane~\cite{Movva2017}. Instead, we can determine $E_Z$ by measuring the sequence of inter-LL energy gaps, $\Delta_\nu$. When $E_Z$ is larger than the cyclotron energy ($E_N = \frac{e \hbar B}{m^*}$, with $m^*$ the effective mass), as expected in the VB of WSe$_2$, the LL sequence separates into two distinct regimes, illustrated in Fig.~3a. LLs of opposite isospin are filled alternately at high filling factor, whereas only one polarization of the isospin is accessible at low filling. We refer to these regimes as ``mixed'' and ``polarized'', respectively. In the polarized regime, the inter-LL energy gaps are equal to the cyclotron energy ($E_N = \Delta_\nu$), in contrast with the mixed regime where two consecutive gaps add up to the cyclotron energy ($E_N = \Delta_\nu + \Delta_{\nu+1}$). Hence, we expect an abrupt change in $\Delta_\nu$ at the transition between the two regimes. The single-particle model predicts $E_Z/E_N \approx 2.2$ in the VB, which means that the first two gaps reside in the polarized regime. The existence of a polarized regime in a single-particle model is unique for WSe$_2$, with typical 2D systems having $E_Z/E_N \ll 1$. 

\begin{figure*}
\centering
\includegraphics[width=17cm]{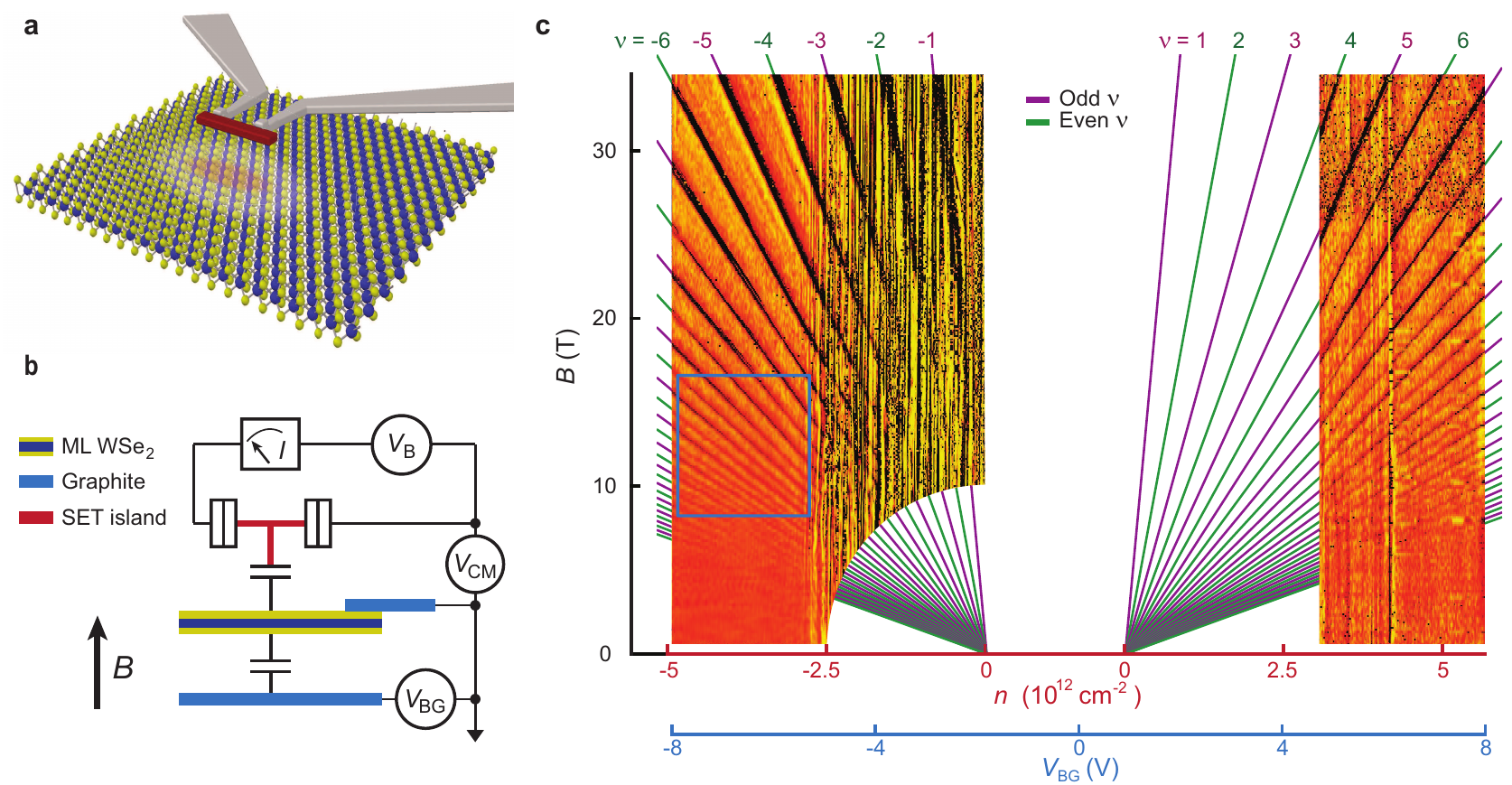} 
\caption{\textbf{| Probing scheme and ambipolar Landau level dispersion.} \textbf{a}, Illustration and \textbf{b}, schematic of the experimental setup. The SET is located closely above the WSe$_2$ and interacts electrostatically with a sub-micrometer area underneath its island electrode (red). A thin ($\sim$30 nm) layer of BN (not shown) separates the SET from the WSe$_2$. We detect changes in the local chemical potential $\mu$ and inverse electronic compressibility $d\mu/dn$ of the WSe$_2$ by measuring the SET current $I$ in response to a common-mode bias voltage $V_{CM}$ (see Supplementary Information). The charge carrier density $n$ in the WSe$_2$ is tuned with a global back gate voltage $V_{BG}$. \textbf{c}, Inverse compressibility $d\mu/dn$ versus charge density $n$ and magnetic field $B$, displaying an ambipolar Landau fan. A background has been subtracted from the data and regions dominated by noise and spurious features are left out for clarity (see Supplementary Information).}
\end{figure*}

Our probing technique directly measures variations in $\mu$, which allows us to accurately determine $\Delta_\nu$. An example of $\mu$ versus $B$ for fixed charge density is shown in Fig.~3b, and the extracted gaps are plotted in Fig.~3c. At high filling (low $B$, \emph{i.e.} in the mixed regime) we find that $\Delta_\nu$ increases linearly with $B$ for both odd and even $\nu$. The pairwise sum of these gaps gives $E_N$, shown as black squares. It is clear from this plot that $\Delta_{-5}$ deviates strongly from the other $\Delta_{\mathrm{odd}}$ and instead falls in line with $E_N$, as expected for gaps in the polarized regime. This holds true over the whole range of $n$, and also for $\nu = -4$, as shown in Fig.~3d. Here, the cyclotron energies for all charge densities in the VB are collapsed into one plot, using the bare values of $\Delta_{-4}$ and $\Delta_{-5}$ and summing the rest pairwise. Disorder complicates the measurement near the band edge, so we restrict our analysis to $n < -2.5\times 10^{12}$ cm$^{-2}$. In our second sample, we can also measure $\Delta_{-3}$ and find that it follows the trend of the cyclotron gaps (see Supplementary Information). We thus conclude that ML WSe$_2$ indeed has an isospin-polarized regime, as predicted by the single-particle model. However, our observation of 5 polarized gaps rather than 2 suggests a significantly larger than expected effective g-factor.

\begin{figure*}
\centering
\includegraphics[width=17cm]{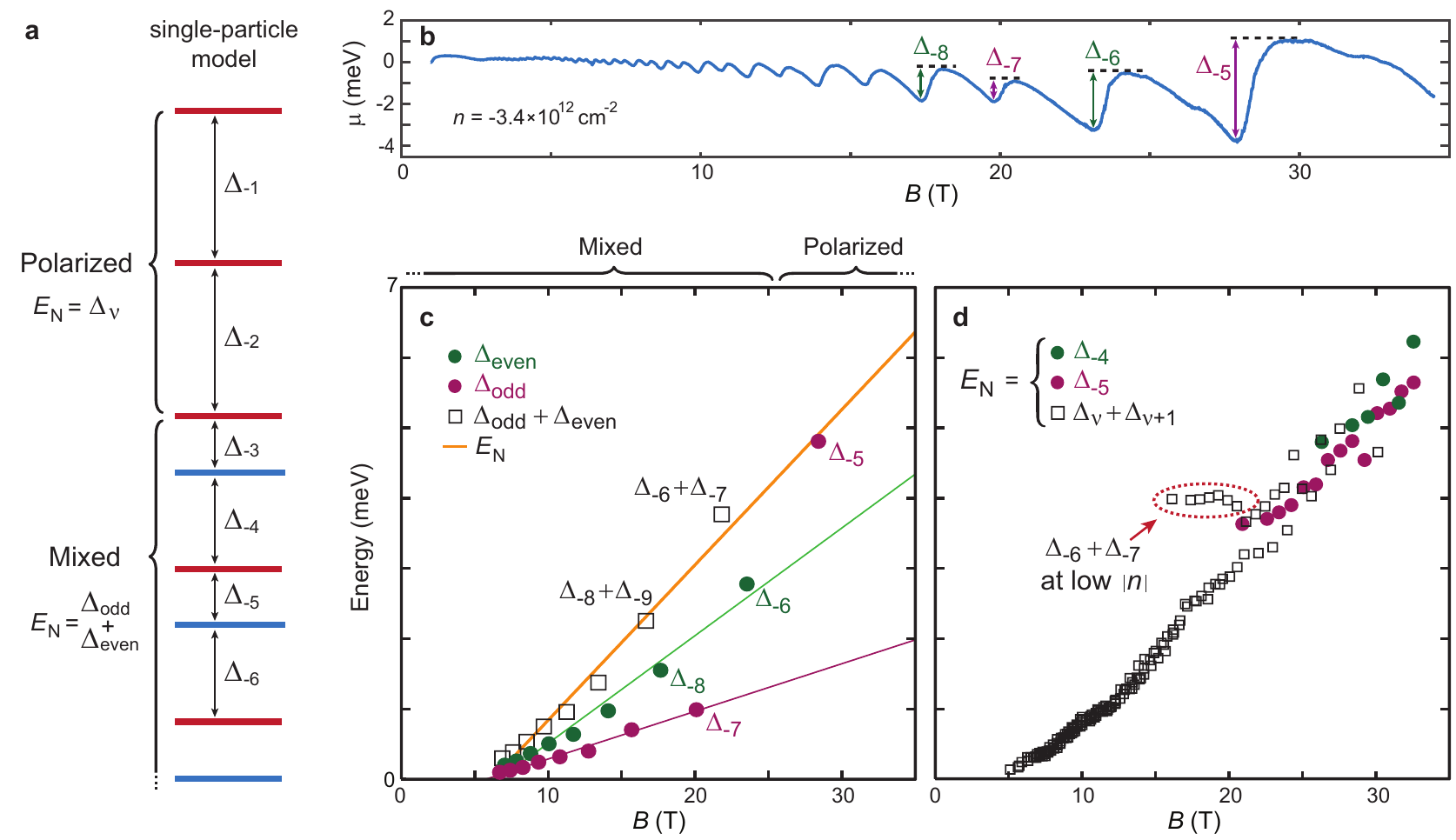}
\caption{\textbf{| Isospin polarization of Landau levels.} \textbf{a}, Illustration of the LL structure in the VB, as predicted by a single-particle model. Since $E_Z > E_N$, the sequence of LL gaps separates into a valley/spin-polarized regime and a mixed regime where the LL filling alternates between the two valleys (red and blue, respectively). In the polarized regime, each energy gap equals the cyclotron energy ($E_N = \Delta_\nu$), whereas in the mixed regime, the cyclotron energy is given by the sum of two subsequent energy gaps ($E_N = \Delta_\nu + \Delta_{\nu+1}$). In the single-particle model of the VB shown here, the polarized regime is expected to encompass two LL gaps. \textbf{b}, Chemical potential $\mu$ (arbitrary offset) and extraction of inter-LL energy gaps $\Delta_{\nu}$, here for a fixed density of $n = -3.4\times10^{12}$~cm$^{-2}$. \textbf{c}, Extracted gaps $\Delta_\nu$ for even (green) and odd (purple) $\nu$ from panel \textbf{b}. $\Delta_{-5}$ deviates strongly from the other $\Delta_{\mathrm{odd}}$ and instead falls in line with the cyclotron gaps $E_N$ of the mixed regime, which implies that it belongs to the mixed regime. We extract the effective mass $m^*$ by fitting a line through $E_N$ versus $B$ (orange) using data from both regimes. The density-dependent polarization beyond the 5 fully polarized gaps, is determined from fits to $\Delta_{\mathrm{even}}$ (green line). \textbf{d}, The equivalent of \textbf{c} but with all gaps in the VB included, emphasizing that $\Delta_{-4}$ and $\Delta_{-5}$ reside in the polarized regime. $\Delta_{-6}$ and $\Delta_{-7}$ belong to the mixed regime but grow anomalously large at very low hole density ($|n| < 3.3\times 10^{12}$ cm$^{-2}$, dashed oval), a feature not expected in any single-particle model but possibly arising due to exchange-induced LL repulsion at the onset of the mixed regime.}
\end{figure*}

In the mixed regime, the different magnitudes of the even and odd gaps manifest as different slopes of the green and purple lines in Fig.~3c. Fig.~4a shows a zoom-in on the portion of the VB highlighted in Fig.~2c by a blue box. Notably, we observe an evolution of the dominant gap parity from even at low hole density to odd at high hole density, using the width of the spikes in $d \mu/dn$ as a proxy for the magnitude of the gaps. This implies that the relative offset between the two isospin LL energies is density dependent -- a feature which is not predicted in a single-particle model with parabolic bands (see Supplementary Information). This observation is borne out more concretely in plots of $\mu$ versus $B$ at different hole densities (Fig.~4b), which demonstrate a clear change in the dominant parity. Plotting the gaps extracted at fixed $B$ = 6 T (Fig.~4c), we find that $\Delta_{\mathrm{odd}}$ and $\Delta_{\mathrm{even}}$ evolve oppositely with $n$, with a cross-over point at $n\approx -3.8\times 10^{12}$ cm$^{-2}$, independent of $B$ (white dotted line in Fig.~4a).

\begin{figure}
\centering
\includegraphics[width=8.8cm]{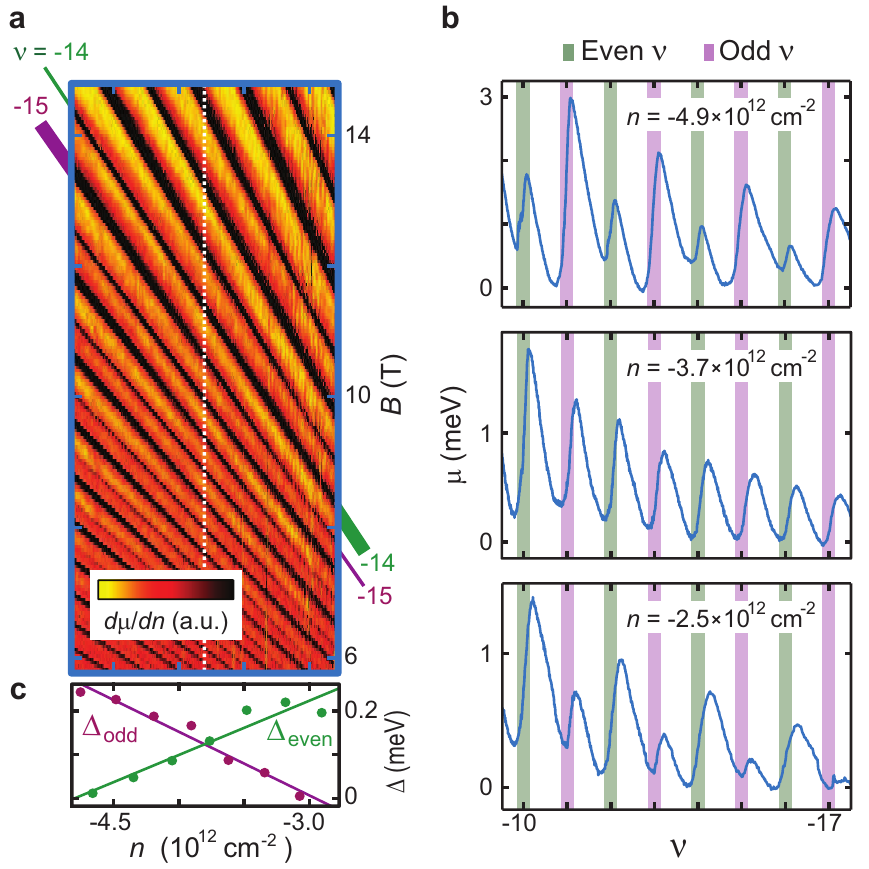} 
\caption{\textbf{| Density dependent Landau level energy gaps in the valence band.} \textbf{a}, Zoom-in on the valence band (blue outline in Fig.~2a), where the LL gap magnitude alternates between high and low (wider, darker lines correspond to larger gaps). The dominant gaps occur at odd $\nu$ for high hole density and gradually shifts to dominance for even $\nu$ at low density. The crossover occurs at $n\approx -3.8\times 10^{12}$ cm$^{-2}$ (dashed white line), and has no discernible dependence on $B$. \textbf{b}, Chemical potential $\mu$ acquired along three lines at fixed $n$ in the VB, plotted against filling factor $\nu$. At high hole density (top), gaps at odd filling dominate, whereas the ones as even filling dominate at low density (bottom). Near the cross-over density (middle), consecutive gaps are equal, apart from a monotonic dependence on $B$. \textbf{c}, Energy gaps between LLs in the valence band for $B$ = 6 T, color-coded by the parity of $\nu$ (odd $\nu$ in purple, even in green).}
\end{figure}

The transition point between the mixed and polarized regime bounds the Zeeman splitting to $5~E_N<E_Z<6~E_N$ in the VB. We determine the remaining fractional part from the ratio of $\Delta_{\mathrm{even}}$ to $E_N$ for each density (green and orange lines in Fig.~3c). The resulting $E_Z/E_N$ is plotted against density in Fig.~5a. To further quantify the density-dependent electronic structure of the material, we extract $m^*(n)$ from the linear fits to $E_N$ versus $B$ (orange line in Fig.~3c) and plot it in Fig.~5b. We find $m^* \gtrsim$ 0.5, which is notably higher than typical predictions by \textit{ab initio} and tight-binding model calculations (green bar)~\cite{Shi2013,Chang2014,Zibouche2014,Kormanyos2015,Fang2015,Wang2017}. Once $m^*(n)$ is known, we can extract $g^*(n) = \frac{E_Z}{E_N} \frac{m_0}{m^*}$. We find that $g^*$ varies between $\sim$10 and $\sim$11.7 with decreasing hole density over the measurement range (Fig.~5c), and is on average enhanced by approximately a factor of 2 compared to the prediction from the single-particle model (dashed line). Fig.~5d illustrates the LL structure predicted by the single-particle model, while Fig.~5e shows the actual structure inferred from our data.

\begin{figure*}
\centering
\includegraphics[width=17cm]{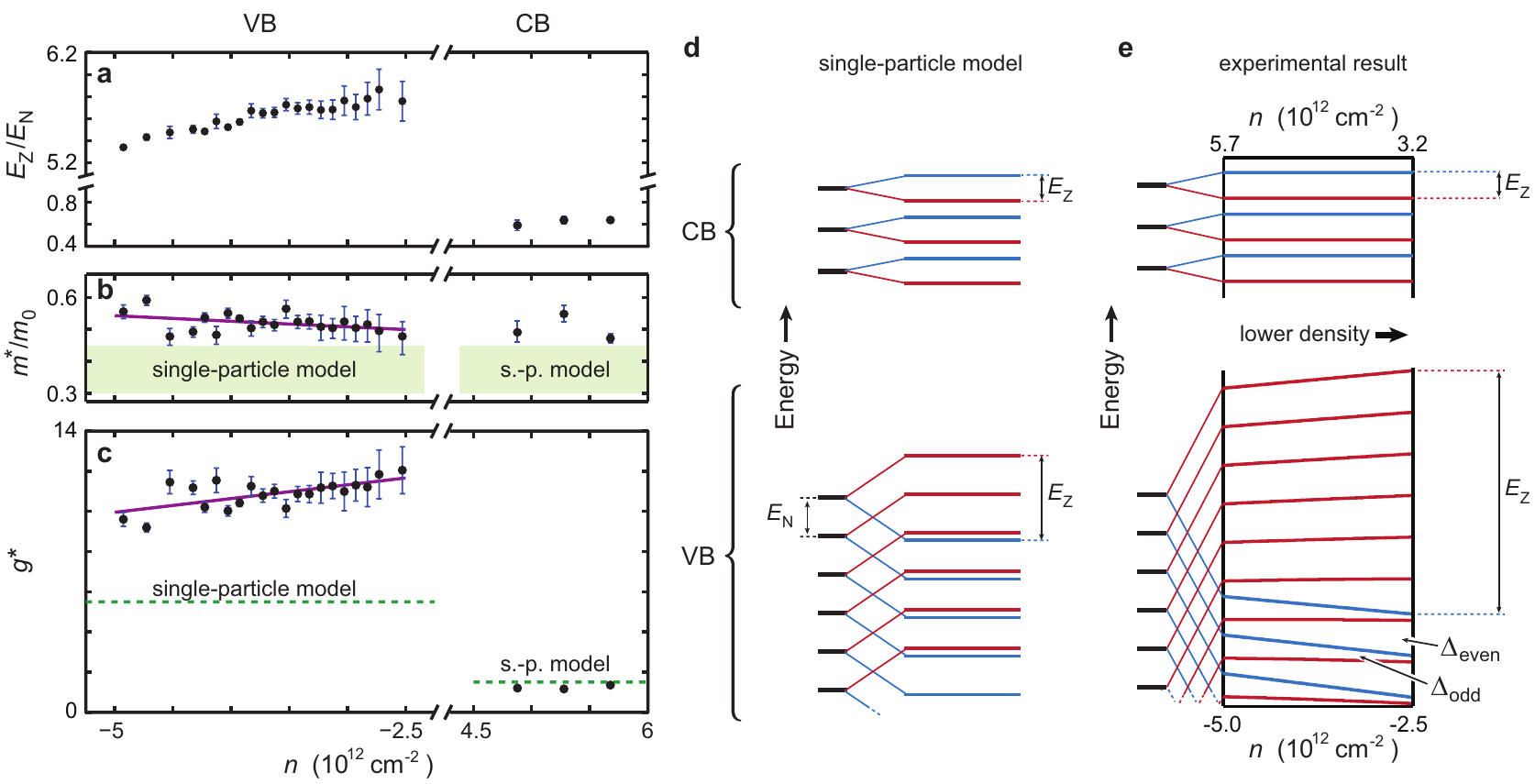}
\caption{\textbf{| Extracted parameters and effects of interactions.} 
\textbf{a}, $E_Z/E_N$ for each accessible density in the VB and CB. \textbf{b}, Effective carrier mass and \textbf{c}, Land\'e g-factor extracted from the data (black dots), and best-fit lines (purple). The green area and dotted line represent typical theoretical predictions from a single-particle model. Both $m^*$ and $g^*$ are enhanced in the VB compared to such predictions. The extraction of $g^*$ in the CB assumes that there is no polarized regime (see text and Supplementary Information).  Error bars in all plots represent the standard error in the slope of the linear fits shown in Fig.~3c. \textbf{d}, Zeeman splitting of the LLs in the single-particle model and \textbf{e}, in the presence of interactions, taking the $m^*$ and $g^*$ shown by solid purple lines in \textbf{b} and \textbf{c} for the VB and average values of the three points in the CB. The offset between the spin-locked valleys in the VB is given by a fixed integral polarization of $5\,E_N$ and a residual density-dependent contribution. In the CB, we observe an odd-dominant sequence of gaps at all accessible electron densities, but our experiment cannot distinguish between the plotted sequence and one with the opposite shift between isospins.}
\label{fig:model}
\end{figure*}

In summary, we observe an unexpectedly high numbers of polarized gaps in the VB, correspondingly large values of $m^*$ and $g^*$, and a dependence of $E_Z/E_N$ on charge density, effects which can all be understood within the context of strong many-body interactions. When electrons occupy many Landau levels, the energy scale of Coulomb interactions is given by $E_C = \frac{e^2}{\epsilon R_C} = \frac{e^3 B}{\epsilon \hbar \sqrt{2 \pi n}}$, where $R_C$ is the cyclotron radius and $\epsilon$ the electric permittivity~\cite{Goerbig2011}. This is several times higher than the cyclotron energy $E_N$ even at the high carrier densities we probe, which suggests \textit{a priori} that many-body interactions play an important role in defining the electronic structure of the material. Coulomb interactions are less strongly screened at lower charge density, scaling as $n^{-1/2}$, and enhancements of $m^*$ and $g^*$ can be expected to follow similar trends, increasing with decreasing density. Strong exchange interactions in particular can significantly enhance $g^*$ and thereby $E_Z$, and thus give rise to the high number of polarized LLs we observe in the VB. The resulting dependence of $E_Z/E_N$ on density explains the evolution from an odd- to even-dominant gap sequence~\cite{Movva2017}, as has previously been observed in low-density Si, AlAs, and GaAs quantum wells~\cite{Kravchenko2000,Vakili2004,Tan2006}. Carrier interactions may also explain the unexpectedly large $\Delta_{-6}$ and $\Delta_{-7}$ at low density (Fig.~3d), consistent between our two samples, as an exchange-driven LL repulsion~\cite{Ando1974} when the system transitions from the polarized to the mixed regime. Such an anomaly does not have a natural explanation in the single-particle model. We note that we may alternatively consider $E_C \propto \sqrt{B}$, which is expected to be appropriate in the quantum limit where all electrons reside in the lowest LL~\cite{Goerbig2011}. In this case the Coulomb and cyclotron energies scale differently with $B$, which should lead to a $B$ dependence in the parity of the gap sequence in addition to the $n$ dependence. However, we do not observe such a dependence, as evidenced by the fact that $\Delta_{\mathrm{odd}}$ and $\Delta_{\mathrm{even}}$ are each linear with respect to $B$ in Fig.~3c. This trend holds across all densities, implying $E_C \propto B$ within our experimental regime.

Finally, we consider the case of the CB, in which the single-particle model predicts a smaller $g^* \approx$ 1.5, because the $E_{Z,B}$ and $E_{Z,S}$ terms have opposite signs and $E_{Z,O} \approx 0$. As a result, contrary to the VB, we expect no polarized regime in the CB (see schematic in Fig.~1b). All experimentally observable gaps in the CB ($\nu \geq$ 4) appear to belong to the mixed regime from inspection of Fig.~2c, suggesting that the Zeeman energy is indeed smaller in the CB than in the VB. Moreover, we see in Fig.~2c that the LL gaps remain odd-dominant across the entire accessible electron-doped regime, also in contrast to the behavior of the VB. Together, this suggests that the role of many-body effects is less prominent in the CB than in the VB -- possibly owing to the smaller initial Zeeman scale -- and highlights the highly asymmetric properties of the two bands.

Our extraction of the ambipolar LL structure of WSe$_2$ demonstrates the important effects of the unusually strong Zeeman energy combined with strong many-body enhancement even at high carrier densities, with enhancements of $E_Z/E_N$ up to a factor of 2.6 in the VB. In conjunction with the high density of states, this suggests the possibility that exchange interactions are strong enough to satisfy the Stoner criterion, implying potential itinerant ferromagnetism at zero magnetic field. This property, which has only recently been observed in a select few 2D materials~\cite{Gong2017,Huang2017}, would additionally be field effect tunable in the case of ML WSe$_2$. The large decoupling of the different isospin components of the LL orbital wave functions could also lead to unusual competitions between fractional quantum Hall and charge density wave ground states at high magnetic field. 

\subsection*{Methods}
The WSe$_2$ is grown in the form of bulk crystals, using a custom chalcogen flux method. Tungsten powder (99.999\%)  and selenium pellets (99.999\%) are loaded in a fused quartz ampoule in the appropriate ratios, with quartz wool acting as a filter for decanting, and sealed under vacuum, $\sim 1\times10^{-3}$ Torr. Subsequently, the ampoule is heated to 1000~$^\circ$C over 24 hours and held at this temperature for 2 days. The ampoule is then slowly cooled to 450~$^\circ$C and centrifuged. Afterwards, the crystals are removed from the ampoule, placed in another ampoule under vacuum and annealed above the melting point of selenium for 48 hours with a 100~$^\circ$C gradient to remove any excess selenium. The difference in data quality between the home-grown WSe$_2$ and commercial material is illustrated Supplementary Fig. 4. 

We use Scotch tape to exfoliate monolayers from the WSe$_2$ crystals and verify their thickness and quality by Atomic Force Microscopy (AFM). By the same method, we produce thicker flakes of hexagonal boron nitride (hBN) and graphite from crystalline material. We use a dry-transfer technique to assemble a multilayer stack where the WSe$_2$ is contacted with graphite flakes, embedded in hexagonal boron nitride, and placed on a graphite backgate. Electrical contact to the graphite sheets are made by standard e-beam lithography and metal deposition (gold with a 3 nm thick sticking layer of titanium underneath).

The SETs are fabricated directly on the upper layer of hBN, selectively in locations where the material is clean as determined by AFM. The patterning is done by e-beam lithography in two-layer resist (MMA copolymer and CSAR 62). The SETs are made of aluminum, deposited by two-angle evaporation with intermediate oxidation in dry O$_2$ gas to form the tunnel barriers.

\subsection*{Acknowledgements}
We thank Kin Fai Mak, Andrea Young, Ben Feldman, and Mark Goerbig for valuable technical and theoretical discussions. CRD and JH acknowledge support from the US Department of Energy, DE-SC0016703. CRD acknowledges partial support from the David and Lucille Packard foundation. Sample fabrication and materials synthesis was supported by the NSF MRSEC program through Columbia in the Center for Precision Assembly of Superstratic and Superatomic Solids (DMR-1420634). K.W. and T.T. acknowledge support from the Elemental Strategy Initiative conducted by the MEXT, Japan and JSPS KAKENHI Grant Numbers JP15K21722. A portion of this work was performed at the National High Magnetic Field Laboratory, which is supported by National Science Foundation Cooperative Agreement No. DMR-0654118, the State of Florida and the U.S. Department of Energy.

\subsection*{Author contributions}
M.V.G. and M.Y. performed the experiments and analyzed the data. M.Y., M.V.G. and C.R.D. developed the model of the system and wrote the paper. M.Y., M.V.G., and C.F. fabricated the samples. D.R. grew and characterized the single crystal WSe$_2$, and K.W. and T.T. grew the single crystal hBN. J.H., X.Z. and C.R.D. advised on experiments, data analysis and contributed to the manuscript writing.

\subsection*{Competing financial interests}
The authors declare no competing financial interests.

\section*{Supplementary materials}
\setcounter{figure}{0}
\renewcommand{\thefigure}{S\arabic{figure}}
\renewcommand{\thesubsection}{S\arabic{subsection}}

\subsection{Zeeman terms in monolayer WSe$_2$}
As a consequence of the spin/valley locking, the Zeeman shift in ML WSe$_2$ has three contributions~\cite{Xu2014}. One is the usual coupling of magnetic field to electron spin, $E_{Z,S} = g_s s \mu_B B$, and the remaining two come from magnetic coupling to the valley-dependent Berry curvature, $E_{Z,B} = \frac{m_0}{m^*} \tau \mu_B B$, and the orbital angular momenta from the tungsten nuclei, $E_{Z,O} = m \tau \mu_B B$~\cite{Aivazian2015}. Here, $\mu_B$ is the Bohr magneton, $g_s$ the spin g-factor, $m$ the orbital magnetic moment, $m_0$ the bare electron mass, $m^*$ the effective carrier mass, and $s = \pm 1/2$ and $\tau = \pm 1$ are the respective quantum numbers for spin and valley. $E_{Z,B}$ and $E_{Z,O}$ arise as direct consequences of the broken inversion symmetry of the lattice, and the Zeeman shift of each compound isospin is the sum of the three terms. The total Zeeman energy splitting between isospins is thus $E_Z = 2\,|E_{Z,S} + E_{Z,B} + E_{Z,O}| = 2\,g^*\mu_B B$, where $g^*$ is an effective Land\'e g-factor. The lowest spin-split bands primarily consist of tungsten $d$-orbitals with $m \approx \pm 2$ in the valence band (VB) and $m \approx 0$ in the conduction band (CB)~\cite{Komider2013}. $s$ and $\tau$ have the same sign in the VB and opposite signs in the CB ~\cite{Kormanyos}which, combined with the difference in $m$, makes the Zeeman shift more pronounced in the VB (Fig.~1b).

\subsection{Energy gaps versus density}

\begin{figure}[h]
\centering
\includegraphics[width=12cm]{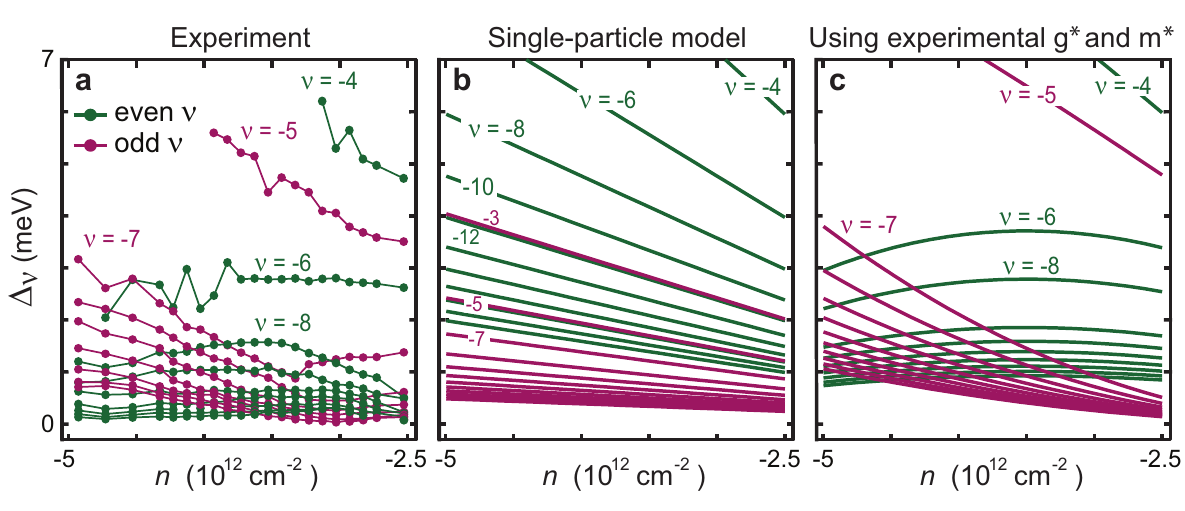} 
\caption{\textbf{| Energy gaps versus density.} \textbf{a}, Inter-LL energy gaps $\Delta_\nu$ extracted from the  VB, grouped by filling factor $\nu$ and color coded by the parity of $\nu$. \textbf{b} Corresponding single-particle calculation and \textbf{c}, calculation using the experimentally determined $g^*(n)$ and $m^*(n)$.}
\label{fig:gaps_vs_n}
\end{figure}

\fref{gaps_vs_n}a shows all accessible $\Delta_\nu$ in the VB, plotted against density. The gaps for $\nu \le -6$ follow an alternating trend consistent with the mixed regime, but the gaps for $\nu \geq -5$ are much larger and increase sequentially, implying that they belong to the polarized regime. The data are in disagreement with the single particle model (\fref{gaps_vs_n}b), which predicts two polarized gaps, but are well reproduced by a calculation using experimentally determined values for $m^*(n)$ and $g^*(n)$ (\fref{gaps_vs_n}c).

\subsection{Data acquisition and processing}
The SET consists of a small metallic island coupled to source and drain electrodes by tunnel junctions. At fixed source-drain bias voltage $V_B$, the current $I$ through the SET is periodically modulated by the charge induced on the island, with a period of the electron charge $e$. In our samples, the only relevant gate capacitance is the one between the SET island and the WSe$_2$ layer. Any change in the local chemical potential of the WSe$_2$ is equivalent to a gate voltage acting through the geometric part of the SET-WSe$_2$ capacitance, $C_{geom}$, thus influencing the SET current.

\begin{figure}[h]
\centering
\includegraphics[width=12cm]{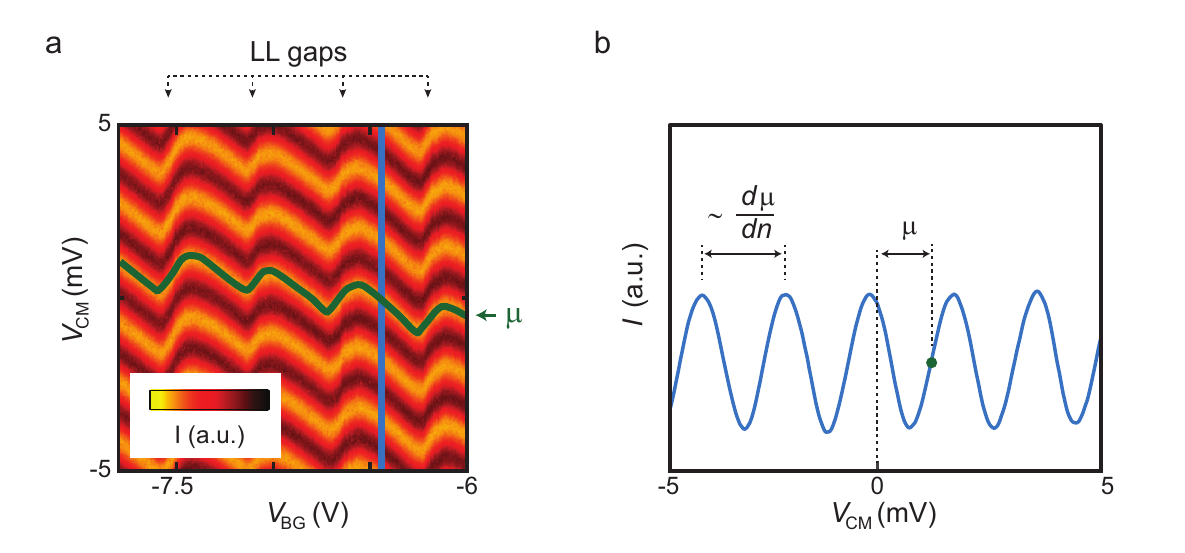} 
\caption{\textbf{| Raw SET data.} \textbf{a}, Example of raw data from the SET, where the color scale shows the SET current $I$. This data set shows a partial sweep of $V_{BG}$ at fixed $B$. Sweeps of $B$ at fixed $V_{BG}$ look similar and are processed in an analogous way. \textbf{b}, By fitting a sine function to each trace of $I$ versus $V_{CM}$, we can extract changes in $\mu$ and $d\mu/d\,n$ with backgate voltage or (similarly) with magnetic field.}
\label{fig:measurement}
\end{figure}

Apart from the fixed bias voltage $V_B$ across the SET, we can apply a common-mode voltage $V_{CM}$ to both source and drain, relative to the WSe$_2$ contact (Fig.~2b). As we sweep $V_{CM}$ over a small range, the SET has a gating effect on the WSe$_2$, thus slightly altering its local chemical potential. As a result, the effective SET-WSe$_2$ capacitance, $C_{SET}$ contains a contribution related to the density of states of the WSe$_2$, given by $1/C_{SET} = 1/C_{geom} + d\mu/d\,n$. Thus, we can use the modulation period of the SET with respect to $V_{CM}$ to determine how $d\mu/d\,n$ changes with $B$ and $V_{BG}$. In a typical experiment, we rapidly and repeatedly sweep $V_{CM}$ over 5-10 SET modulation periods while slowly varying $V_{BG}$ or $B$. An example of the raw data from such a measurement is shown in \fref{measurement}a. The $I$ versus $V_{CM}$ curve acquired for each value of $V_{BG}$ or $B$ can be fitted with a function of the form $I\propto \sin\;\frac{2 \pi C_{SET}}{e} (V_{CM}  + \frac{\mu}{e})$. We extract $\mu$ and $d\mu/d\,n$ from such fits, as illustrated in \fref{measurement}b.

\begin{figure}[h]
\centering
\includegraphics[width=12cm]{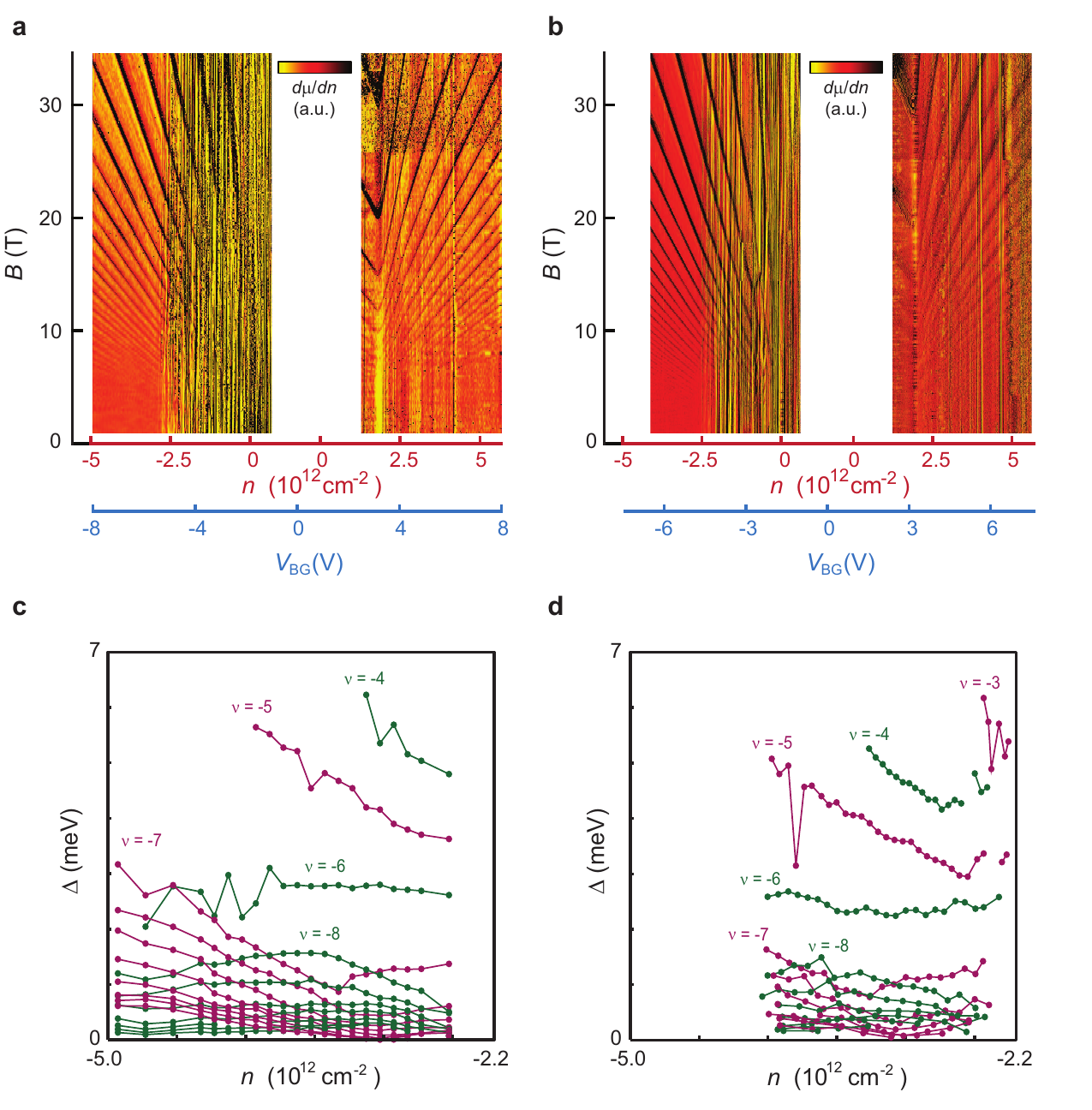} 
\caption{\textbf{| Sample comparison.} LL fans for \textbf{a}, sample A, \emph{i.e.} the one presented in the main text. \textbf{b}, The equivalent fan for sample B, made with WSe$_2$ from a different growth cycle. \textbf{c}, LL gaps $\Delta_\nu$ for sample A and \textbf{d}, sample B. In sample B, we can access the the $\nu=-3$ gap and confirm that it resides in the polarized regime.}
\label{fig:sample_comparison}
\end{figure}

The data in Figs. 2c, 4a, and 4c were acquired by slowly varying $V_{BG}$ at fixed $B$, as were all data acquired on sample B (Figs. \ref{fig:sample_comparison}b and d). Before extracting gaps from such data sets (\fref{sample_comparison}d and Fig. 4c), we subtract a background to compensate for electrostatic effects unrelated to the LL structure. The background is taken as the mean of all $\mu$ versus $I$ curves over a wide range of $B$, so that features related to the LLs are averaged away. The data in all other figures were acquired by sweeping $B$ at fixed $V_{BG}$. In this case we do not subtract any background before extracting gaps. To improve the clarity of the features in Figs. 2c and 4a, we take the average of $d\mu/d\,n$ across all values of $B$ for each value of $V_{BG}$ and subtract this from the data. In all data sets, we have removed abrupt jumps due to large random changes in the SET charge offset.

\begin{figure}[h]
\centering
\includegraphics[width=12cm]{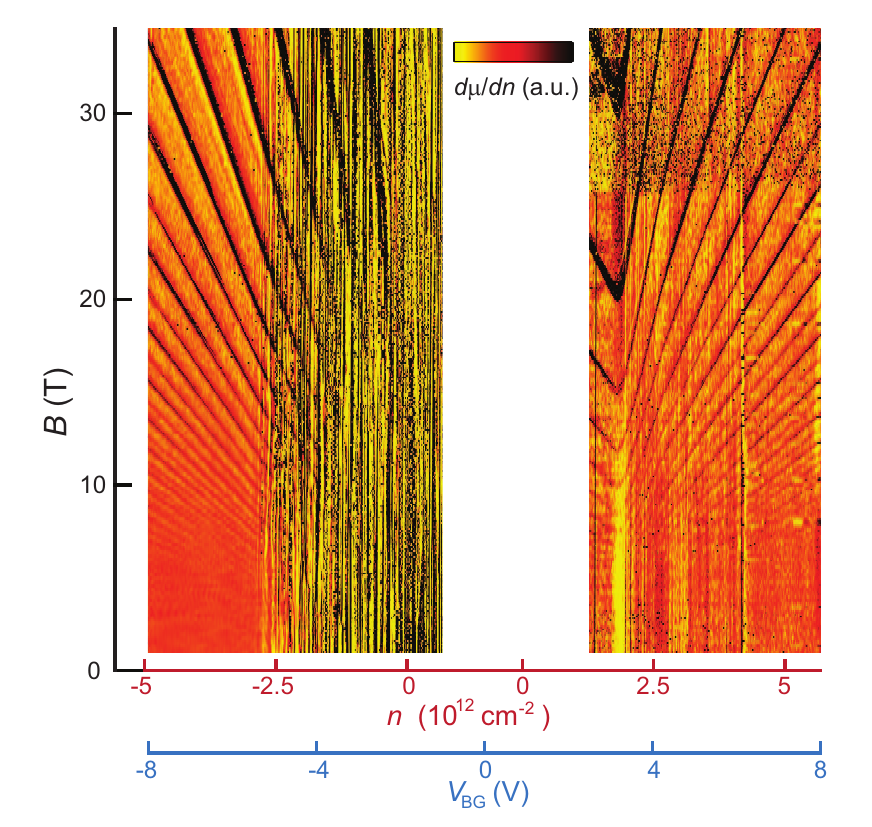} 
\caption{\textbf{| Complete LL data.} Full data set corresponding to Fig. 2c. The same background subtraction as in Fig. 2c has been applied.}
\label{fig:full_fan}
\end{figure}

In Fig. 2c, we have excluded regions of the parameter space which are irrelevant due to noise and spurious features. The full data set is shown in \fref{full_fan}. The curved features at the lowest density in the CB are not repeatable between measurements and appear to depend on the sweep rate of $V_{BG}$ as well as on where the sweep begins in relation to the CB edge. We attribute them to Schottky effects at the contact to the WSe$_2$.

The data presented in the main text were all acquired on the same sample (A), but we have also performed experiments on a second sample (B, \fref{sample_comparison}). We find that the results are qualitatively very similar, including the anomalous rise in $\Delta_{-6}$ and $\Delta_{-7}$ at low hole density. We can also confirm from sample B that $\Delta_{-3}$ is consistent with the polarized regime. In sample A, the gaps were acquired from sweeps of $B$ at fixed $n$, while in sample B they were acquired from sweeps of $n$ at fixed $B$. \fref{material_comparison}b shows a third sample using commercially available WSe$_2$, exhibiting significantly higher disorder than the home-grown samples (\fref{material_comparison}a).

\begin{figure}[h]
\centering
\includegraphics[width=12cm]{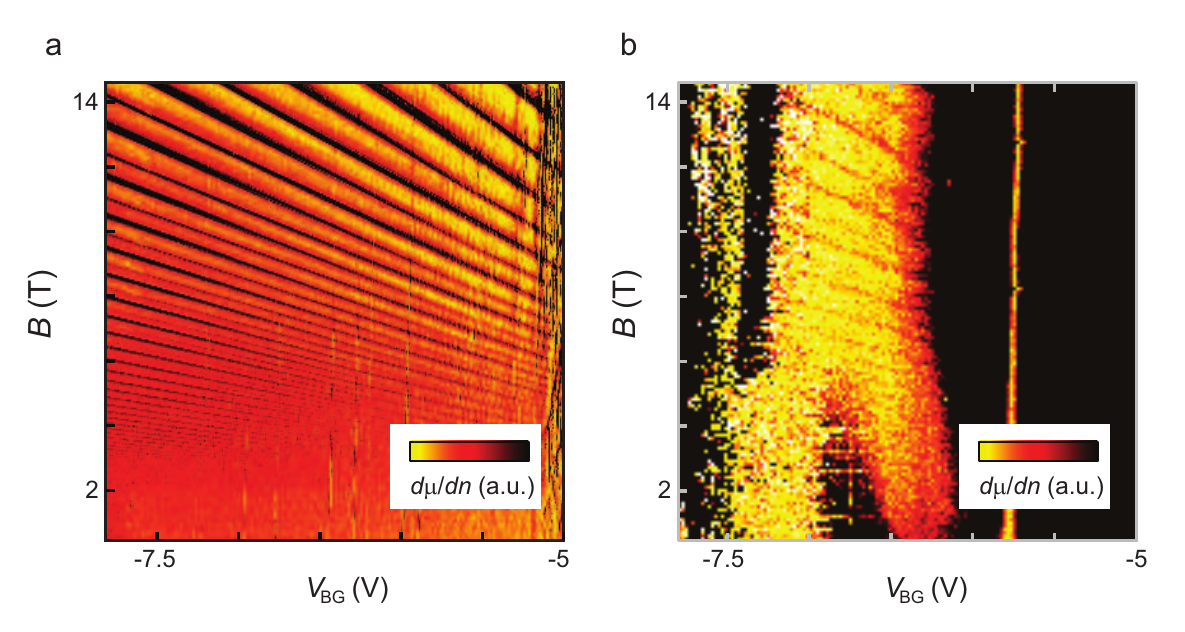} 
\caption{\textbf{| Material comparison.} Data for ML samples using \textbf{a}, homegrown and \textbf{b}, commercial WSe$_2$. No background has been subtracted. The two data sets were acquired at different resolutions and with different instrument settings, but nonetheless serve to illustrate the difference in material quality.}
\label{fig:material_comparison}
\end{figure}

\subsection{Effect of the material band gap}

By projecting the LL gaps in each band to zero magnetic field, we find that the gate voltage needs to change by $\Delta V_{BG} = 2.7$ V in order to traverse the bandgap of the material, $E_G$. In a system where the back gate provides the only relevant capacitance to the WSe$_2$, $d\mu/d\,V_{BG} = e$ in the gap, so $E_G = \Delta V_{BG}$. In our case, however, the presence of the SET counteracts some of the doping from the back gate. With the quantum capacitance defined by $\frac{1}{C_Q} = \frac{d \mu}{d\,n}$, we have

$$\frac{d\mu}{d V_{BG}} = \frac{e}{1+\frac{C_{SET} + C_Q}{C_{BG}}}\;,$$where capacitances are given per unit area. In a parallel-plate approximation of the SET, $C_{SET}/C_{BG}\approx 0.6$, which results in a bandgap of $E_G \approx \frac{\Delta V_{BG}}{1.6} = 1.7$ eV. However, this value is uncertain due to the geometry of the SET and the resulting distortion of the electric field.

In principle, mid-gap defects result in non-zero $C_Q$ within the gap. However, we believe this to be a negligible effect in our samples, as our estimated defect density from STM imaging is $<5 \times 10^{10}$ cm$^{-2}$. Assuming each defect can be singly charged, this would at most lead to an adjustment of $\sim$50 meV.

\subsection{Identification and details of the conduction band valley}

In addition to the valleys at the corners of the first Brillouin zone (K/K' points), semiconducting TMDs have low-energy valleys at other symmetry points, notably at the center of the zone ($\Gamma$) and at a point roughly halfway between $\Gamma$ and K, referred to as Q~\cite{Kormanyos2015}. In the VB of monolayer WSe$_2$, the valley at K is unambiguously the lowest energy band, and the spin-splitting has been shown to be large ($\sim$500 meV). Therefore, we clearly are probing the higher spin-split VB at the K valleys experimentally. The situation in the CB is less clear, since the valleys at Q and K are nearly degenerate and the spin-splitting at K is roughly an order of magnitude smaller than in the VB~\cite{Kormanyos2015}. Experimentally, we find that we always probe the lowest spin-split band at the K valleys. We can rule out contributions from the Q valleys because they should be 3-fold degenerate, as the entire Q-point valley lies inside the first Brillouin zone, while our LL structure clearly shows 2-fold degeneracy as evidenced by the even-odd parity structure. We can then calculate the anticipated electron doping necessary to reach the higher spin-split band in the K valleys assuming a parabolic band dispersion with $m^* = 0.52m_0$, and find this to be slightly larger than is experimentally accessible. This is confirmed by our lack of observation of a second set of LLs dispersing from high electron density. These arguments justify our modeling in the main text, which considers only the lower spin-split valleys at the K points in both the VB and CB.

As discussed in the main text, the inability to probe gaps at small electron density prevents an assignment of the number of polarized gaps in the CB. However, examining $d\mu/dn$ in Fig.~\ref{fig:full_fan} suggests that the alternating, odd-dominant sequence of gaps persists down to at least $\nu$ = 3, suggesting $P \leq$ 2. Most likely, however there is no polarized regime due to the small starting $g^*$ in the single-particle model. Under this assumption, there are two possible LL orderings , where either of the two compound isospins could be the ground state. Furthermore, both of these correspond to $g^* \approx$ 1 in our model, hence they are completely indistinguishable experimentally without a probe of the magnetization. The effect of interactions on $g^*$ is unusual in the CB due to the competing spin and Berry Zeeman contributions: as the spin term grows, $g^*$ must shrink to zero before increasing again. Assuming a similar amount of enhancement of $g_s$ in the CB as we observe in the VB, the ground state would be the case where the isospin energies flip due to interactions. However, it may also be the case that the lack of a polarized regime leads to smaller exchange enhancement, hence the single-particle model may also remain the correct description.

\subsection{Effects of band anharmonicity}

To leading order, \textit{ab initio} and tight-binding model calculations predict parabolic dispersions at the VB and CB edges at the K-points in monolayer WSe$_2$. In this case, the cyclotron energy and all Zeeman terms scale linearly with $B$, and the relative LL hierarchy between locked isospins is therefore independent of $B$. The leading order correction to the massive Dirac fermion Hamiltonian used to describe the band edges is to include second order expansion terms in crystal momentum $k$. Following Ref.~\cite{Rostami2013}, the cyclotron energy including these $k^2$ terms is given by

\begin{equation} \label{eq:lls}
E_N = \pm \sqrt{\bigg[\frac{\Delta - \lambda \tau s}{2} + \frac{e \hbar B}{2 m_0} \Big(\beta N - \frac{\alpha \tau}{2}\Big)\bigg]^2 +\frac{2 e N B (t_0 a_0)^2}{\hbar}} + \frac{\lambda \tau s}{2} +  \frac{e \hbar B}{2 m_0} \Big(\alpha N - \frac{\beta \tau}{2}\Big),
\end{equation}

where $\Delta$ is the energy gap and $\lambda$ is the spin-orbit coupling. As these parameters are not all well known, we take values similar to those found from previous \textit{ab initio} calculations, tuned slightly to match the anticipated isospin splitting with parabolic bands for simplicity. As such, we use $\Delta$ = 2.7 eV, $\lambda$ = 0.4 eV, $t_0$ = 1.05 eV, $\alpha$ = 0.43, $\beta$ = 2.21, and $a_0$ = 0.332 nm, though our primary conclusions here do not depend at all on the exact values chosen. As the cyclotron energy now has terms that scale with both $B$ and $\sqrt{B}$, while the Zeeman energy scales linearly with $B$, the curvature of the LLs may in principle result in even-to-odd transitions of the dominant gap even in the absence of many-body interactions. 

\begin{figure}[h]
\centering
\includegraphics[width=12cm]{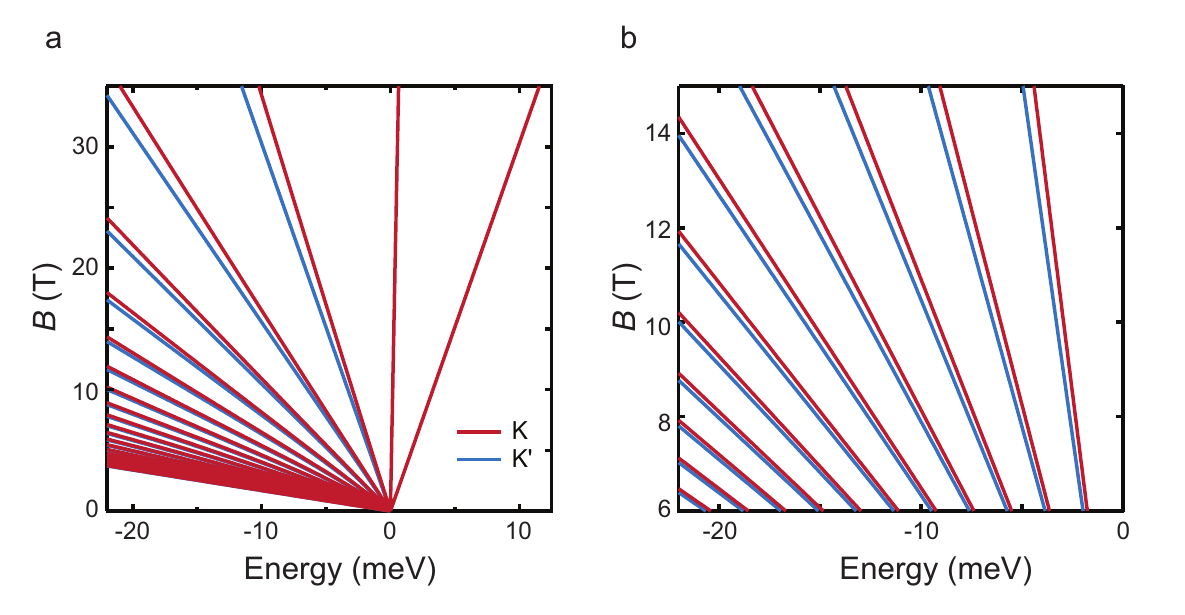}
\caption{\textbf{| Effects of band anharmonicity.} LL dispersion with anharmonic corrections to the band structure. \textbf{a}, The LL structure of the VB resembles the prediction considering parabolic bands over the experimentally relevant range of energy and $B$. \textbf{b}, Zooming in to the level structure more clearly illustrates that the LL curvature is a small effect.}
\label{fig:anharmonicity}
\end{figure}

To test whether this effect is potentially of large enough magnitude to explain our observations, we calculate the LL structure using Eq.~\ref{eq:lls} along with the Zeeman terms described in the main text. \fref{anharmonicity}a plots the first few LLs in the VB for each valley over an energy range comparable to our experimental doping range, showing a similar structure to the prediction in Fig.~1b of the main text. \fref{anharmonicity}b shows a zoom-in over a smaller range of $B$, where it becomes more clear that the degree of LL curvature does not result in a significant density or field dependence of the relative gap size of each parity. Quantitatively the effect is much less than 5\%, while we observe the relative gap size change by well over 50\% with density in our experiment. This strongly suggests that the parity effect we observe is due to many-body interactions, and cannot be understood in a single-particle model.

\end{document}